\documentclass[conference]{IEEEtran}
\IEEEoverridecommandlockouts
\usepackage{amsmath,amssymb,amsfonts}
\usepackage{tabularx}
\usepackage{booktabs, tabularx}
\usepackage{multirow}
\usepackage{array}
\usepackage[caption=false,font=normalsize,labelfont=sf,textfont=sf]{subfig}
\usepackage{textcomp}
\usepackage{stfloats}
\usepackage{url}
\usepackage{verbatim}
\usepackage{graphicx}
\usepackage{cite}
\usepackage{pifont}
\usepackage{lettrine}
\usepackage{mathtools}
\usepackage{hyperref}
\usepackage[export]{adjustbox}
\usepackage{caption}
\usepackage{subcaption}
\usepackage[ruled, lined, linesnumbered, commentsnumbered, longend]{algorithm2e}
\newcolumntype{M}[1]{>{\centering\arraybackslash}m{#1}}
\newcolumntype{N}{@{}m{0pt}@{}}
\usepackage{makecell}

\usepackage[top=54pt, left=48pt, right=48pt, bottom=69pt]{geometry}

\newcommand{\squeezeup}{\vspace{-2mm}}

\begin{document}
	
	\title{Deep Learning-Based Data Fusion of 6G Sensing and Inertial Information for Target Positioning: Experimental Validation}
	
	\makeatletter
    \newcommand{\linebreakand}{%
    \end{@IEEEauthorhalign}
    \hfill\mbox{}\par
    \mbox{}\hfill\begin{@IEEEauthorhalign}
}
\makeatother
	
	\author{
    \IEEEauthorblockN{
        Karthik Muthineni\IEEEauthorrefmark{1}\IEEEauthorrefmark{3}, Alexander Artemenko\IEEEauthorrefmark{1}, Artjom Grudnitsky\IEEEauthorrefmark{2},  Josep Vidal\IEEEauthorrefmark{3}, Montse Nájar\IEEEauthorrefmark{3} 
    }
    \IEEEauthorblockA{\IEEEauthorrefmark{1} Corporate Sector Research and Advance Engineering, Robert Bosch GmbH, Renningen, Germany}
    \IEEEauthorblockA{\IEEEauthorrefmark{2} Nokia Bell Labs, Stuttgart, Germany}
    \IEEEauthorblockA{\IEEEauthorrefmark{3} Department of Signal Theory and Communications, Universitat Politècnica de Catalunya (UPC), Barcelona, Spain}
    \IEEEauthorblockA{Email: \IEEEauthorrefmark{1}$\{${karthik.muthineni, alexander.artemenko}$\}$@de.bosch.com, \IEEEauthorrefmark{3}$\{${josep.vidal, montse.najar}$\}$@upc.edu}
}
	
	
		
	
	\maketitle
	
\begin{abstract}
The sixth-generation (6G) cellular technology will be deployed with a key feature of Integrated Sensing and Communication (ISAC), allowing the cellular network to map the environment through radar sensing on top of providing communication services. In this regard, the entire network can be considered as a sensor with a broader Field of View (FoV) of the environment, assisting in both the positioning of active and detection of passive targets. On the other hand, the non-3GPP sensors available on the target can provide additional information specific to the target that can be beneficially combined with ISAC sensing information to enhance the overall achievable positioning accuracy. In this paper, we first study the performance of the ISAC system in terms of its achievable accuracy in positioning the mobile target in an indoor scenario. Second, we study the performance gain achieved in the ISAC positioning accuracy after fusing the information from the target's non-3GPP sensors. To this end, we propose a novel data fusion solution based on the deep learning framework to fuse the information from ISAC and non-3GPP sensors.   

We validate our proposed data fusion and positioning solution with a real-world ISAC Proof-of-Concept (PoC) as the wireless infrastructure, an Automated Guided Vehicle (AGV) as the target, and the Inertial Measurement Unit (IMU) sensor on the target as the non-3GPP sensor. The experimental results show that our proposed solution achieves an average positioning error of $3~\textrm{cm}$, outperforming the considered baselines. 
		
\end{abstract}
	

\section{Introduction}
The upcoming sixth-generation (6G) mobile communication technology is expected to extend its footprint from the traditional consumer market to vertical industries by promising to deliver advanced functionalities that open new use cases \cite{BOSCH}. The 6G technology envisions taking the network capability beyond communications by enabling ``radar-like" sensing to detect objects and their positions without having them equipped with a radio transceiver. This is achieved with a functionality called Integrated Sensing and Communication (ISAC), with integration performed at different stages, such as spectrum sharing between radar sensing and communication functions, use of the same hardware, and reuse of the communication waveform and radio resources for sensing \cite{Zhang2022, Wei2023}. Nevertheless, this integration increases the system's complexity and opens new challenges. The development of cellular networks relied heavily on foundation models and theories. However, these approaches face limitations due to increased network complexity as we move from generation to generation. Recent years have witnessed the enormous capabilities of Artificial Intelligence (AI) algorithms in solving complex models and optimization problems in various applications \cite{BOSCHMOBILITY, BOSCHRESEARCH}. The role of AI can also be witnessed across a broad range of communication topics like signal detection \cite{He2020}, channel encoding \cite{Be2020}, channel estimation \cite{Luo2020}, and resource allocation \cite{Liang2020}. Moreover, the future 6G cellular network is expected to natively support AI processing by adding suitable hardware components into the network design \cite{NOKIA}. Therefore, AI will be key in addressing certain challenges of 6G and the ISAC system \cite{Demirhan2023}.

\begin{figure}[t!]
	\centering
	\includegraphics[clip, trim=4.7cm 6.5cm 6cm 5.5cm, width=0.5\textwidth]{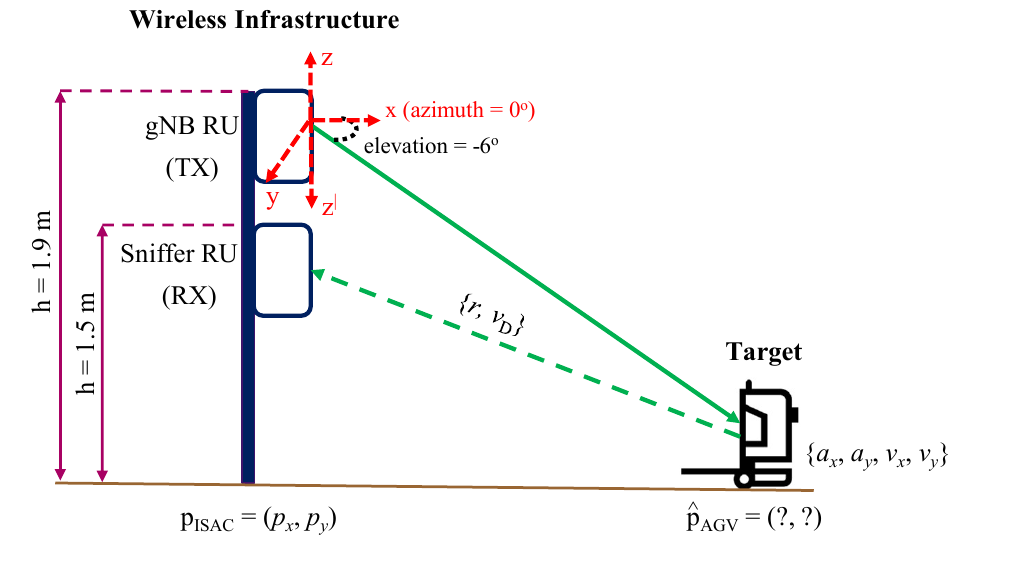}%
	\caption{Block diagram of our ISAC sensing/positioning setup.}
	\label{fig: systemmodel}
        \squeezeup
        \squeezeup
        \squeezeup
\end{figure}

The standardization body 3GPP has conducted a feasibility study on ISAC in Release $19$ and identified $32$ potential use cases \cite{3GPP}. This paper evaluates the sensing/positioning performance of the ISAC system introduced in \cite{Wild2023} and the potential of AI models in enhancing the ISAC-assisted positioning accuracy for one of the 3GPP's use cases, i.e., tracking Automated Guided Vehicles (AGVs) in complex scenarios. Earlier works on ISAC-assisted tracking focused on evaluating the ISAC system performance in simulated environments. In \cite{Favarelli2022}, authors considered multiple mono-static radars for fusing sensing information corresponding to a single target and compared the tracking performance against different algorithms. A joint framework for estimating the target location and its clock drift in the 6G ISAC system was presented in \cite{WEIJUNDA2023}. In \cite{VIJAYA2023}, the authors proposed using an AI model for positioning the target by fusing the sensing information collected by multi-static radars in an indoor scenario. 

There is still a research gap in ISAC literature investigating how to fuse ISAC sensing information with the data obtained from other sensor modalities, such as the non-3GPP sensors available on the target. We contribute to ISAC literature by (\textit{i}) evaluating the achievable positioning accuracy of the real mobile target in an indoor scenario using a Proof-of-Concept (PoC) of ISAC, (\textit{ii}) demonstrating the value of non-3GPP sensors and their provided information in enhancing the ISAC-assisted positioning accuracy by fusing the information from these two systems, (\textit{iii}) we propose a novel data-driven Deep Neural Network (DNN) based data fusion approach for fusing the non-3GPP data with ISAC sensing information. Furthermore, we validate our approach through real-world experiments with an ISAC PoC as the wireless infrastructure, the AGV serving as the passive target for positioning, and an Inertial Measurement Unit (IMU) sensor on the target as the non-3GPP sensor. To the best of our knowledge, this is the first study focusing on enhancing ISAC-assisted target positioning accuracy with non-3GPP data and demonstrating the results through real-world experiments.   

\section{Background}
The ISAC PoC used in this work is built upon the 5G communications hardware \cite{Wild2023}. In this section, we present the details of ISAC PoC, the achievable positioning accuracy with ISAC, and the problem under investigation.

\subsection{ISAC System}
We consider a quasi-monostatic\footnote{A configuration where transmitter and receiver are co-located but are slightly separated in distance.} ISAC PoC, which uses base station-centric sensing, i.e., transmission and reception of sensing-relevant signals, and their processing does not involve User Equipment (UE). Sensing is performed using the Orthogonal Frequency Division Multiplexing (OFDM) radar algorithm~\cite{Braun2014}, where a transmitted OFDM radio frame with $\mathrm{G}$ symbols and $\mathrm{H}$ subcarriers, forming a matrix $\textbf{U}$, is reflected by the environment. The reflections are received by the ISAC antennas as an OFDM radio frame forming the matrix $\textbf{V}$, allowing computation of the Channel State Information (CSI) matrix $\textbf{Z}$ by element-wise division of $\textbf{V}$ by $\textbf{U}$. The OFDM radar algorithm then performs $\mathrm{H}$ Fast Fourier Transforms (FFTs) of length $\mathrm{G}$ and $\mathrm{G}$ Inverse Fast Fourier Transforms (IFFTs) of length $\mathrm{H}$ on the CSI matrix $\textbf{Z}$, to obtain a range-doppler matrix~\cite{Braun2010}. We enhance the OFDM radar algorithm to improve target detection with the clutter removal approach described in~\cite{MARCUS2023}. The magnitude of each element in the range-doppler matrix is used to differentiate between noise and potential targets, with a simple threshold-based approach used in the following experiments. For each target thus obtained, we extract the parameters range $r$ and Doppler velocity $v_{\textrm{D}}$.

The PoC realizing this system consists of $2$ Radio Units (RUs): a gNodeB (gNB) RU used for transmission of both sensing and communication signals and an additional RU of the same type used only for receiving reflections of the transmitted signal, called ``Sniffer RU". Both RUs are rectangular arrays, each consisting of $2$ sub-arrays with $8\times12$ antenna elements. The RUs are configured for a center frequency of $f_c=27.4~\mathrm{GHz}$ and a bandwidth of $200~\mathrm{MHz}$ with a subcarrier spacing of $\Delta_f=120~\mathrm{KHz}$. Sensing processing is performed by the Sensing Processing Unit (SPU), in the PoC a separate machine that monitors the fronthaul connection to the gNB RU and directly interfaces via a fronthaul link with the Sniffer RU. The SPU receives the downlink IQ samples from monitoring the gNB RU fronthaul and the reflection IQ samples from the Sniffer RU fronthaul. It uses them to generate the matrices $\textbf{U}$ and $\textbf{V}$ and perform OFDM radar processing as described above. For the experiments, the RUs were mounted as shown in Fig.~\ref{fig: systemmodel} with a mechanical elevation of $-6^{\circ}$ down-tilt and azimuth of $0^{\circ}$.

\begin{figure}[t!]
	\centering
	\includegraphics[clip, trim=13cm 20cm 5.5cm 19.8cm, width=0.32\textwidth]{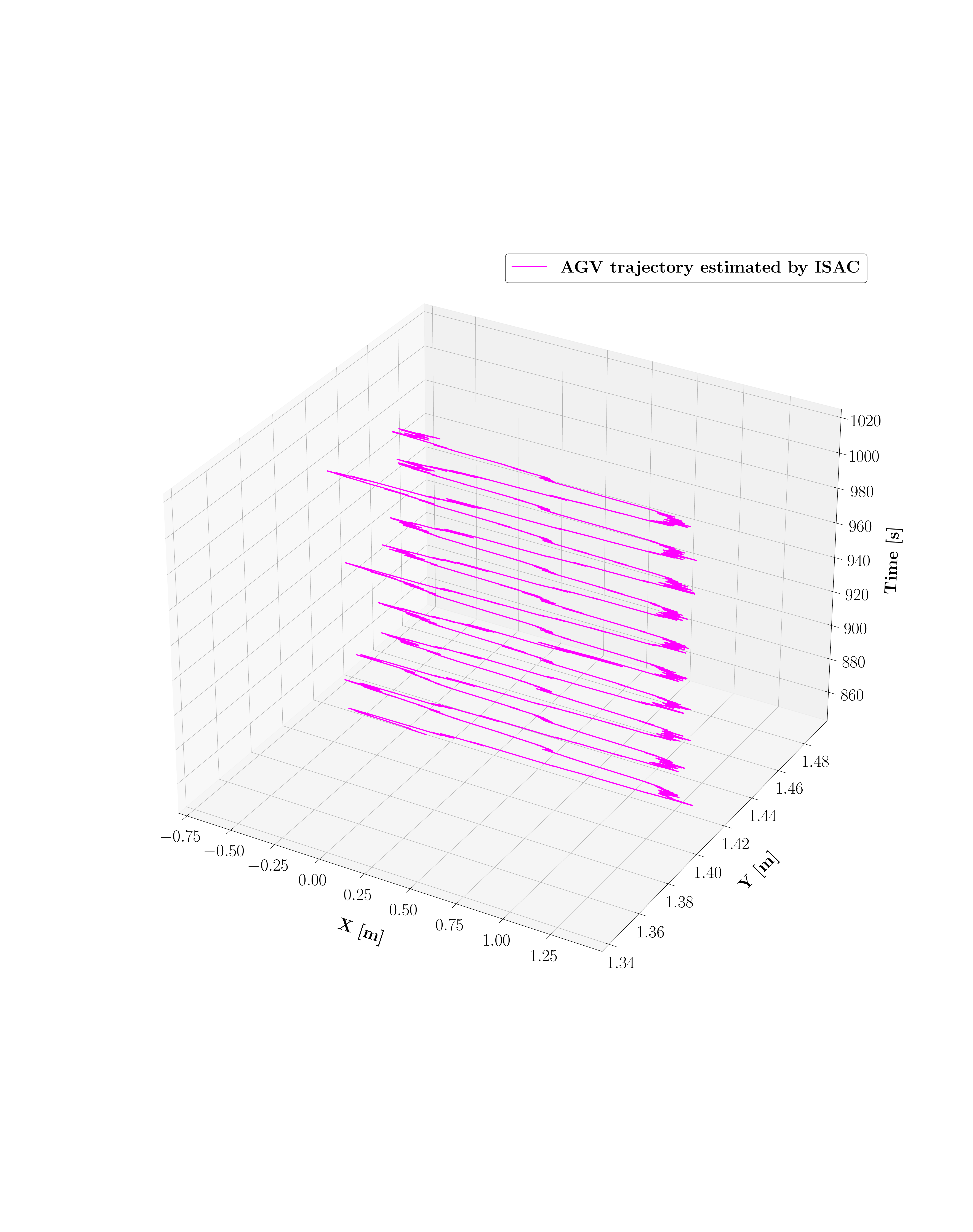}
	\caption{The trajectory of the target AGV estimated by ISAC PoC using the geometric approach (discussed in \ref{subsection: isacpositioning}).}
	\label{fig: ISACestimatedtrajectory}
\end{figure}
\begin{figure}[t!]
	\centering
	\includegraphics[clip, trim=13cm 20cm 5.5cm 19.8cm, width=0.32\textwidth]{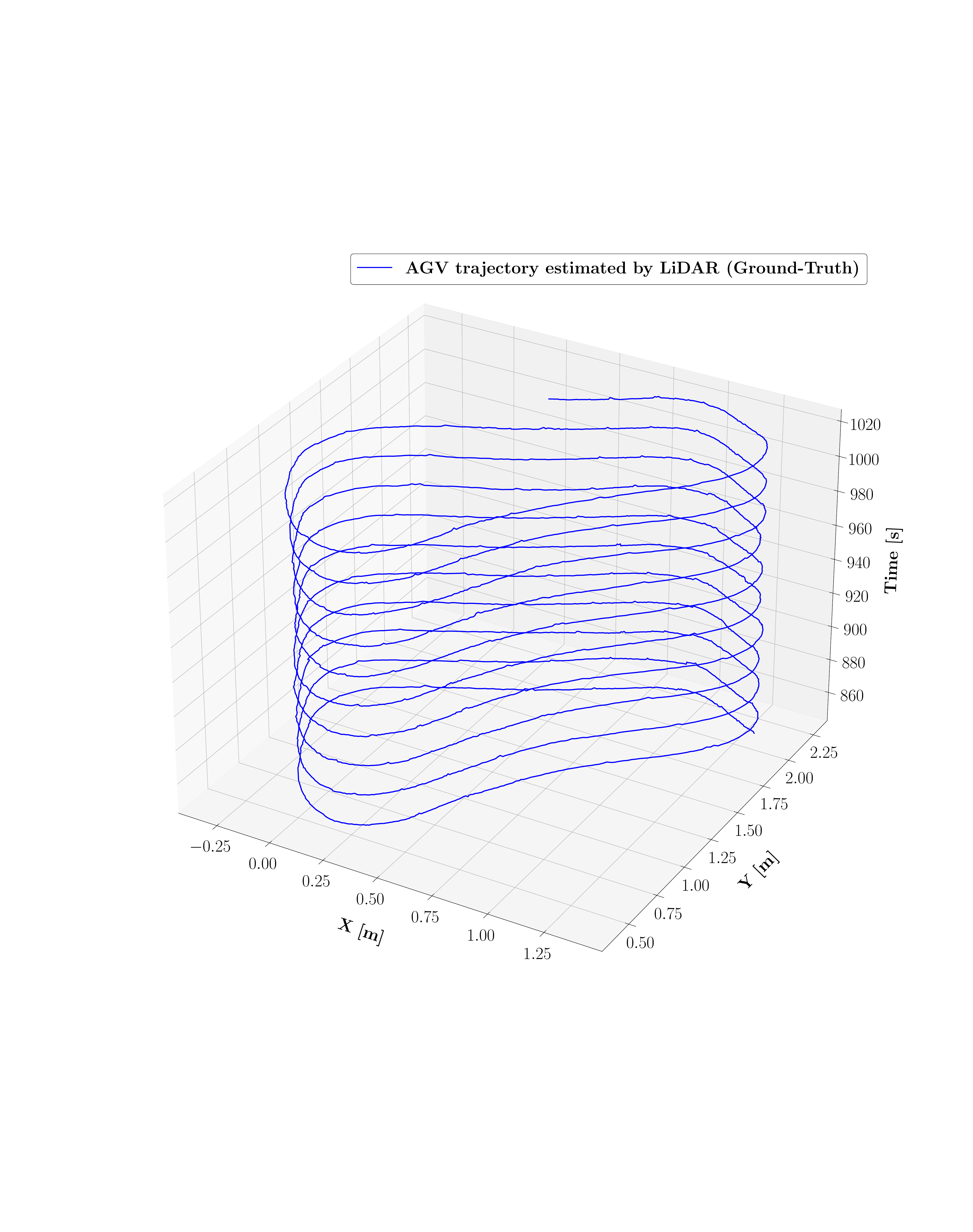}
	\caption{The trajectory of the target AGV estimated by the onboard LiDAR sensor (ground-truth).}
	\label{fig: Lidarestimatedtrajectory}
        \squeezeup
\end{figure}

\subsection{ISAC-Assisted Positioning}
\label{subsection: isacpositioning}
The range estimated by ISAC PoC towards the target corresponds to 3D due to the difference in the heights of the ISAC infrastructure and the target. For 2D positioning (scope of this work), the range in 3D must be projected onto a 2D map for further integration with sensor fusion. This can be achieved with $r_{\textrm{2D}}~=~( (r_{\textrm{3D}})^2 - (\Delta h)^2 )^{\frac{1}{2}}$, where, $\Delta h$ is the difference in the heights between the ISAC infrastructure and the target. Thereafter, the 2D position of the target is estimated using the geometric approach as 

\begin{equation}
    \hat{p}_{x} = p_x + r_{\textrm{2D}}~.~\cos({\Psi}),
    \label{eq1}
\end{equation}
\begin{equation}
    \hat{p}_{y} = p_y + r_{\textrm{2D}}~.~\sin({\Psi}),
    \label{eq2}
\end{equation}
where $(p_x\,,p_y)$ represents the known position coordinates of ISAC PoC and $\Psi$ indicates the azimuth angle of the sniffer RU. Fig.~\ref{fig: ISACestimatedtrajectory} depicts the target trajectory estimated by ISAC PoC. The plot shows that ISAC PoC determines the target to be moving only across the $x$ axis (1D). However, the ground-truth LiDAR sensor on the AGV records the trajectory traversed by the target and is illustrated in Fig.~\ref{fig: Lidarestimatedtrajectory}. Observing both plots, one can infer the limitation of ISAC PoC in positioning the target under this particular investigation scenario. The achievable positioning accuracy with ISAC PoC is found to be $2$~m. A high positioning error is evident because a single wide beam was used to illuminate the target, resulting in receiving echoes reflected from multiple objects. As a result, it becomes difficult to determine the Line-of-Sight (LoS) path delay, leading to inaccurate positioning results. This demands the techniques to complement ISAC-assisted positioning accuracy. To this end, we believe that the non-3GPP sensors can be considered an additional source of information, contributing to enhancing the 3GPP's system (ISAC) performance.  

\begin{figure*}[t!]
	\centering
	\includegraphics[width=0.95\textwidth]{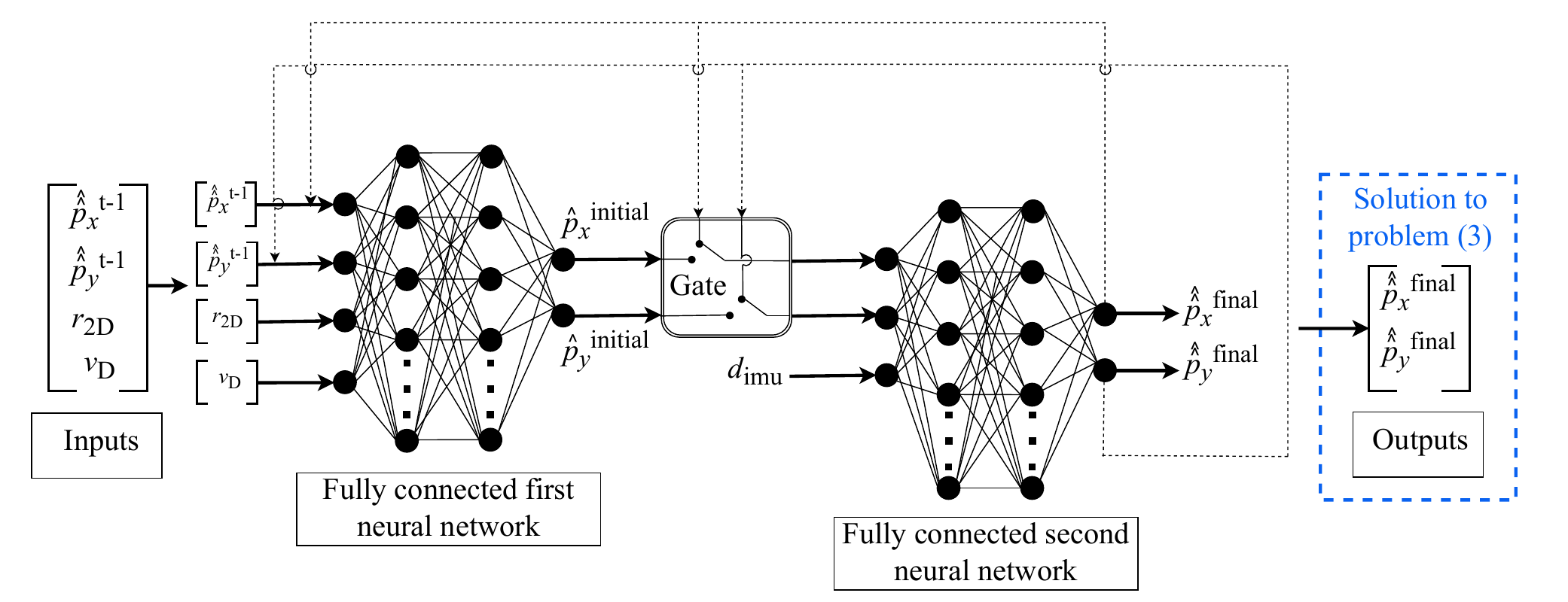}
	\caption{Proposed two-stage cascaded DNN data fusion and positioning solution.}
	\label{fig: cascadeddnn}
        \squeezeup
\end{figure*}

\subsection{Problem Statement}
\label{Problem Statement}
This work considers ISAC PoC as the wireless infrastructure sensor $\textrm{S1}$. The AGV equipped with IMU sensor $\textrm{S2}$ as well as Light Detection and Ranging (LiDAR) sensor $\textrm{S3}$ as the UE and/or non-3GPP sensors on the target to be positioned. The ISAC PoC provides sensing parameters as an output at a frequency of $33$~Hz. The measurement at $i$-th time is represented as  $\Lambda_{i}^{(\textrm{S1})}~=~\{t_i, \textbf{x}_{\textrm{ISAC}}\}$, where, $t_i$ is the timestamp and $\textbf{x}_{\textrm{ISAC}}$ constitutes to the state vector incorporating range and Doppler velocity of a target, $\textbf{x}_{\textrm{ISAC}} = [r, v_{\textrm{D}}]^\mathrm{T}$. The IMU sensor provides its motion parameters as an output at a frequency of $50$~Hz. The measurement at $i$-th time is represented as $\Lambda_{i}^{(\textrm{S2})}~=~\{t_i, \textbf{x}_{\textrm{IMU}}\}$. The state vector of the IMU sensor contains accelerations and angular velocities in $x$, $y$, and $z$ axes, $\textbf{x}_{\textrm{IMU}}~=~[a_x, a_y, a_z, g_x, g_y, g_z]^\mathrm{T}$. In addition, LiDAR is used as a ground-truth sensor, which provides the 2D position of AGV and can be represented as $\Lambda_{i}^{(\textrm{S3})}~=~\{t_i, \textbf{x}_{\textrm{LiDAR}}\}$, $\textbf{x}_{\textrm{LiDAR}}~=~[p^{\textrm{true}}_x, p^{\textrm{true}}_y]^\mathrm{T}$. The problem under investigation is, given the ISAC and the IMU asynchronous measurements, how can the position of AGV be obtained by fusing the sensor measurements? The problem is formulated as

\begin{equation}
    \hat{\textbf{p}}_{\textrm{AGV}} = f\Big(\Lambda^{(\textrm{S1})}, \Lambda^{(\textrm{S2})}\Big),
    \label{eq3}
\end{equation}
where $\hat{\textbf{p}}_{\textrm{AGV}}$~$=$~$[{\hat{p}_{x}}, {\hat{p}_{y}}]^{\mathrm{T}}$ is the estimated 2D position of AGV and $f(\cdot)$ indicates the function that models the relationship between the measurements of the sensors and the AGVs position. In this work, the function $f(\cdot)$ is designed to be a deep learning model.

\begin{algorithm}[t]
\scriptsize
	\SetKwFunction{isOddNumber}{isOddNumber}
	\SetKwInOut{KwIn}{Training phase}
	\KwIn{ }
        Acquire measurements from the wireless infrastructure ISAC $\Lambda^{(\textrm{S1})}$ and the IMU sensor $\Lambda^{(\textrm{S2})}$ through a measurement campaign. \\
        Convert accelerations from the IMU sensor to distance measurement $ d_{\textrm{imu}}$. \\
        Use position estimates created by the LiDAR sensor as ground-truth and train both the DNNs to provide the position of the AGV $\hat{\textbf{p}}_{\textrm{AGV}}$ as the output. \\
\scriptsize
        \SetKwInOut{KwOut}{Testing phase}
	\KwOut{ }
	\For{$i \geq 1$}{For the current time step $t_i$, acquire measurements from the sensors $\Lambda_{t_i}^{(\textrm{S1})}$ and $\Lambda_{t_i}^{(\textrm{S2})}$. \\ 
        \eIf{ISAC data is available at $t_i$}
            {Provide $\left[\hat{\hat{p}}_x^{t-1}, \hat{\hat{p}}_y^{t-1}, r, v_{\textrm{D}}\right]^\mathrm{T}$ as inputs to the first stage of DNN. \\
            Determine the initial position of target AGV $\left[\hat{p}_x, \hat{p}_y \right]^\mathrm{T}$. \\
            Provide $\left[\hat{p}_x, \hat{p}_y \right]^\mathrm{T}$ and $d_{\textrm{imu}}$ as inputs to the second stage of DNN to estimate the target's final position $\left[\hat{\hat{p}}_x, \hat{\hat{p}}_y\right]^\mathrm{T}$.
            }{
            Provide $ \left[\hat{\hat{p}}_x^{t_{i}-1}, \hat{\hat{p}}_y^{t_{i}-1} \right]^\mathrm{T} $ and $d_{\textrm{imu}}$ as inputs to the second stage of DNN to estimate the target's final position.
            }      
        }
\caption{Proposed algorithm for the problem (\ref{eq3})}
\label{alg1}
\end{algorithm}

\section{Proposed DNN-based data fusion architecture}
This section presents the details of the proposed two-stage cascaded DNN data fusion and positioning solution and the approach used for training the network.

\subsection{Two-Stage Cascaded DNN}
Our proposed data fusion architecture is based on deep learning. The DNNs can learn the features from the raw data and are known to perform remarkably well with the labeled datasets~\cite{Zhao2019}. Therefore, we propose a two-stage cascaded DNN to fuse the measurements from ISAC PoC with the IMU sensor data to estimate the 2D position of the target, as shown in Fig.~\ref{fig: cascadeddnn}. The first stage of the DNN takes ISAC measurements as the input and provides the target's initial position as the output, which acts as input to the second stage of the DNN, along with the IMU sensor data. Following the inputs, the second stage of the DNN provides the final estimate of the target position as output, resulting from the fusion between the two sensor modalities. In the first stage of the DNN, the previous final estimated position of the target $\left(\hat{\hat{p}}_x^{t-1}, \hat{\hat{p}}_y^{t-1}\right)$, range $r$, and Doppler velocity $v_{\textrm{D}}$ are the inputs. By incorporating the temporal data $\left(\hat{\hat{p}}_x^{t-1}, \hat{\hat{p}}_y^{t-1}\right)$ as inputs to the DNN, it helps the network to understand the underlying patterns in the data and make predictions that align with the dynamic changes. 

The IMU sensor provides the acceleration measurements relative to its local coordinate frame. These measurements need to be translated to the global coordinate frame using a rotation matrix as

\begin{equation}
\begin{bmatrix}
a_{x} \\
a_{y} 
\end{bmatrix}_{\textrm{g}}
=
\begin{bmatrix}
\cos({\Phi}) & \sin({\Phi}) \\
-\sin({\Phi}) & \cos({\Phi})
\end{bmatrix}
\begin{bmatrix}
a_{x} \\
a_{y} 
\end{bmatrix}_{\textrm{l}},
\label{eq4}
\end{equation}
where the local and global coordinate frames are represented by $[a_x, a_y]_{\textrm{l}}^\mathrm{T}$ and $[a_x, a_y]_{\textrm{g}}^\mathrm{T}$ respectively. Furthermore, $\Phi$ corresponds to the yaw angle obtained from the gyroscope. Instead of providing the global accelerations from IMU as input to the second stage DNN, we translate them into the distance traversed by AGV as $ d_{\textrm{imu}} = v\Delta t + \frac{1}{2}a_{\textrm{g}}(\Delta t)^2$, where $v$ and $\Delta t$ refers to the velocity of the AGV and the time interval between two consecutive IMU measurements. The term $a_{\textrm{g}}$ represents total acceleration given by $a_{\textrm{g}} = { \left( (a_{x})^2 + (a_{y})^2 \right) }^\frac{1}{2}$. Therefore, in the second stage of DNN, the inputs are the estimated position of the target $\left(\hat{p}_x, \hat{p}_y\right)$ and the distance traversed by the target $d_{\textrm{imu}}$. It is to be noted that the measurements from ISAC PoC and the IMU sensor are provided at a frequency of $33$~Hz and $50$~Hz, respectively. This implies that the first stage of DNN can provide the target's initial position as output at a frequency of $33$~Hz. However, the second stage of DNN takes the target estimated position $\left(\hat{p}_x, \hat{p}_y\right)$ and the distance traversed $d_{\textrm{imu}}$ as inputs, which are provided at different frequency rates. Therefore, in those time instances where the ISAC measurements are missing, and the estimated position of the target from the first stage of DNN is not available, we take the previous final estimated position of the target $\left(\hat{\hat{p}}_x^{t-1}, \hat{\hat{p}}_y^{t-1}\right)$ as the input to the second stage of DNN. The workflow of the data fusion architecture is summarized in Algorithm~\ref{alg1}.

\subsection{DNN Training and Testing}
A pre-defined trajectory was created and sent to the target AGV to follow the path. The data from the onboard IMU sensor was recorded as the target continued to traverse along the route. In addition, ISAC PoC transmits its single wide beam toward the target via gNB RU and collects the reflections through sniffer RU to derive the sensing parameters. Later, the data was recorded and collected by ISAC, and the IMU sensors were used to train the two-stage cascaded DNN.

Our data fusion architecture was implemented in Python with the Keras framework. The data collected during the measurement campaign was divided into training ($80\%$) and validation ($20\%$) datasets. We use the simplified architecture consisting of four layers in each DNN to keep the model lightweight yet powerful. The four layers comprise $1$ input layer, $2$ hidden layers, and $1$ output layer. The first DNN consists of $\{$$4$, $32$, $16$, $2$$\}$ neurons in corresponding layers. While, the second DNN consists of $\{$$3$, $32$, $16$, $2$$\}$ neurons in corresponding layers. We use a batch size of $32$, representing the number of training samples processed together before updating the parameters (weights and biases) of the DNN model. We consider the Rectified Linear Unit (ReLU) activation function. The Adaptive Moment Estimation (ADAM) optimizer with a learning rate of $10^{-3}$ is used for training. We choose the loss function as Mean Square Error (MSE) for both neural networks to minimize the positioning error between the estimated target position and the ground-truth, which can be formulated as

\begin{figure}[t!]
	\centering
	\includegraphics[clip, trim=4.5cm 6.5cm 2cm 4cm, width=0.5\textwidth]{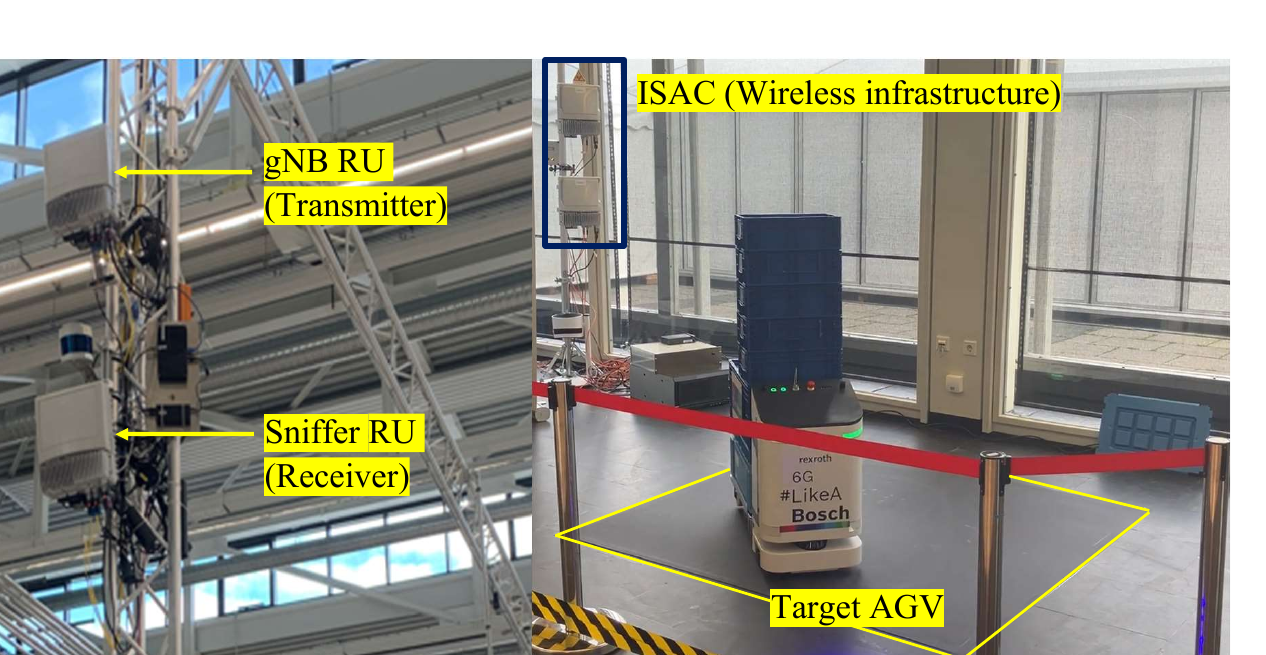}%
	\caption{Experimental setup with the wireless infrastructure ISAC and the target AGV in an indoor scenario.}
	\label{fig: experimentalsetup}
        \squeezeup
        \squeezeup
\end{figure}

\begin{equation}
    \mathcal{L}^{\textrm{MSE}}(\Gamma) = \mathbb{E}\{||\hat{\textbf{p}}_{\textrm{AGV}}-\textbf{p}^{\textrm{true}}_{\textrm{AGV}}||^2_2\},
    \label{eq5}
\end{equation}
where $\hat{\textbf{p}}_{\textrm{AGV}}$ is the estimated position, $\textbf{p}^{\textrm{true}}_{\textrm{AGV}}$ is the ground-truth obtained from the LiDAR sensor of the target AGV, and $\Gamma$ indicates the parameters of the DNN, which include weights and biases. The models of both the DNNs are validated with the validation datasets to prevent overfitting. 

\section{Performance Evaluation}
In this section, we describe the experimental setup and present the analysis results, including the benchmark considered and the accuracy of the obtained positioning. 

\subsection{Experimental Setup}
The experiments were conducted in an indoor scenario with an Area of Interest (AoI) for sensing and/or positioning spanning roughly $3.5$~m~$\times$~$3.5$ m. The ISAC PoC consisting of the gNB RU and the sniffer RU is mounted at the height of $1.9$~m and $1.5$~m, respectively, from the ground level, as shown in Fig.~\ref{fig: experimentalsetup}. Moreover, the RUs are mounted with an azimuth of $0^{\circ}$ and elevation of $-6^{\circ}$ to direct the beam toward the AoI. The AGV from Bosch Rexroth is used as a target and is provided with a pre-defined trajectory to traverse in the AoI. The dimensions of the target correspond to be $1$~m~$\times$~$0.4$~m~$\times$~0.9~m. Furthermore, the target has non-3GPP sensors like IMU and LiDAR (ground-truth).

\begin{figure}[t!]
	\centering
	\includegraphics[clip, trim=15cm 21cm 5.5cm 19.8cm, width=0.32\textwidth]{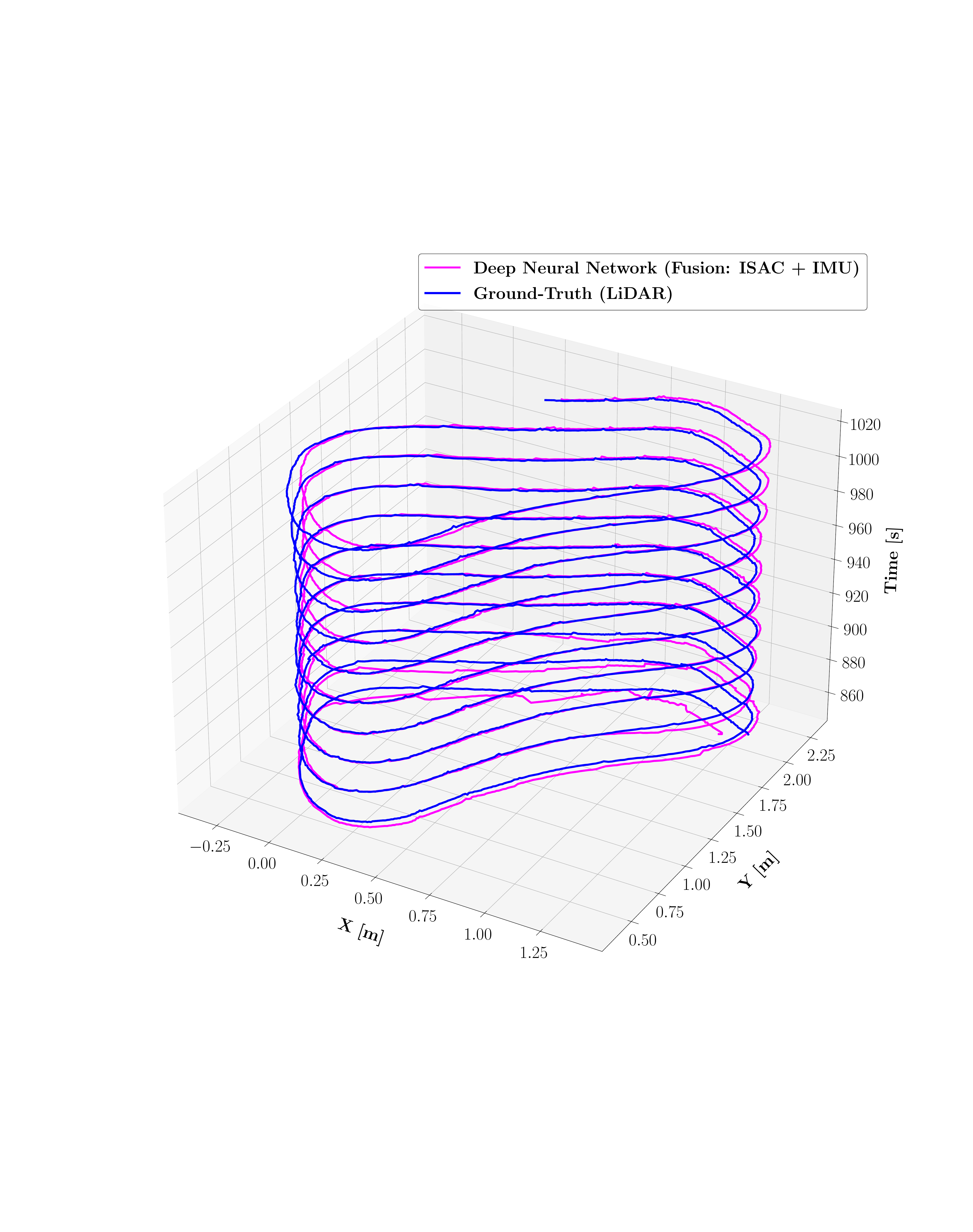}%
	\caption{Ground-truth and the estimated trajectory of the target by the two-stage cascaded DNN.}
	\label{fig: fusion}
        \squeezeup
        \squeezeup
        
\end{figure}

\subsection{Experimental Results}

Using ISAC PoC, we present the positioning results of the target AGV obtained with our proposed two-stage cascaded DNN data fusion framework. Fig.~\ref{fig: fusion} depicts the ground-truth trajectory of the target AGV obtained from the LiDAR sensor and the estimated trajectory computed by fusing the wireless infrastructure ISAC sensing measurements with the data from the onboard IMU sensor of the target AGV. Looking at the trajectory plots, one can infer that our proposed positioning technique estimates the target's trajectory close to the ground-truth and enhances the positioning accuracy compared to the geometric approach (Fig.~\ref{fig: ISACestimatedtrajectory}). The improvement in positioning accuracy shows the value in the information provided by the non-3GPP IMU sensor and also explains how the 3GPP system (ISAC) can benefit from such rich information. Limiting our analysis to only one non-3GPP sensor, IMU, we can achieve an average positioning error of $3~\textrm{cm}$ for the given AoI, corresponding to about $1\%$ relative positioning error. 

Our proposed data fusion positioning technique is compared with two baselines for benchmarking. For the first baseline ``DNN~(ISAC)", we utilize the previous initial estimated position of the target and ISAC sensing parameters $ [\hat{p}_{x}^{t-1}, \hat{p}_{y}^{t-1}, r, v_{\textrm{D}}]^\mathrm{T} $ as the inputs to the first stage of DNN to estimate the target's trajectory without using the second stage of DNN. The second baseline ``EKF~(Fusion:~ISAC~+~IMU)", uses the conventional model-driven approach Extended Kalman Filter (EKF) to fuse ISAC and IMU measurements. To this end, the state vector of the EKF at time $t$ is defined as $\textbf{x}_t~=~[\hat{p}_x, \hat{p}_y, \hat{v}_x, \hat{v}_y]^\mathrm{T}$, where $(\hat{p}_x$\,, $\hat{p}_y)$ and $(\hat{v}_x$, $\hat{v}_y)$ are the position coordinates and velocities of the target in $x$ and $y$ directions, respectively. We use the acceleration measurements from the IMU sensor to predict the state vector, while the ISAC measurements are used to update the state vector. Therefore, the EKF continues to predict the state vector using IMU measurements, and whenever the ISAC measurements become available, the EKF updates the state vector. The motion model is defined as $\hat{\textbf{x}}_{t+1}~=~\textbf{C}\textbf{x}_t~+~\textbf{D}\textbf{u}_t~+~\textbf{w}_t$, where the state transition matrix $\textbf{C}$, the control input matrix $\textbf{D}$, and the control input vector $\textbf{u}$ are defined as
\begin{equation}
    \textbf{C} = \begin{bmatrix}
1 & 0 & \Delta t & 0 \\
0 & 1 & 0 & \Delta t \\
0 & 0 & 1 & 0 \\
0 & 0 & 0 & 1 
\end{bmatrix}, \hspace{2mm}
 \textbf{D} = \begin{bmatrix}
\frac{\Delta t^2}{2} & 0 \\
0 & \frac{\Delta t^2}{2} \\
\Delta t & 0 \\
0 & \Delta t
\end{bmatrix}, \hspace{2mm}
\textbf{u} = \begin{bmatrix}
a_{x} \\
a_{y}
\end{bmatrix}_{\textrm{g}}.
\end{equation}
The time step is represented by $\Delta t$, and the IMU accelerations are used as the control inputs \textbf{u}. In addition, the process noise $\textbf{w}_t$ is considered Gaussian. The error in the state prediction is computed as $\hat{\textbf{E}}_{t+1} = \textbf{C}\textbf{E}_{t}\textbf{C}^\mathrm{T}+\textbf{Q}$, with \textbf{Q} indicating the process noise covariance containing IMU measurement uncertainties. The ISAC PoC provides the range measurements of the target, which can be represented through observation vector as $\textbf{n}_{t+1}~=~||~\textbf{p}_{\textrm{ISAC}}~-~\hat{\textbf{p}}_{\textrm{AGV}}~||~+~m_{t+1}$, with measurement noise $m_{t+1}$. Subsequently, the EKF computes the residual, representing the difference between the actual and predicted measurements $\textbf{o}_{t+1} = \textbf{n}_{t+1}-\textbf{h}(\hat{\textbf{x}}_{t+1})$. The predicted measurements can be obtained by computing Jacobian \textbf{J} as 
\begin{equation}
    \textbf{J}_{t+1} = \frac{\partial \textbf{h}}{\partial \hat{\textbf{x}}_{t+1}} = \begin{bmatrix}
\frac{\hat{p}_x - p_{x}}{||\textbf{p}_{\textrm{ISAC}}-\hat{\textbf{p}}_{\textrm{AGV}}||} & \frac{\hat{p}_y - p_{y}}{||\textbf{p}_{\textrm{ISAC}}-\hat{\textbf{p}}_{\textrm{AGV}}||} & 0 & 0 \\
\end{bmatrix}.
\end{equation}
Given the measurement model, the state vector is updated using $\hat{\textbf{x}}^{+}_{t+1} = \hat{\textbf{x}}_{t+1} + \textbf{K}_{t+1}(\textbf{n}_{t+1}-\textbf{h}(\hat{\textbf{x}}_{t+1}))$, where $\textbf{K}_{t+1}~=~\hat{\textbf{E}}_{t+1}\textbf{J}_{t+1}^\mathrm{T}(\textbf{J}_{t+1}\hat{\textbf{E}}_{t+1}\textbf{J}_{t+1}^\mathrm{T}+\textbf{R})^{-1}$ is the  Kalman gain and \textbf{R} is the measurement covariance, indicating uncertainty in range measurements of ISAC. Thereafter, the error in the state estimation is also updated as $\hat{\textbf{E}}^{+}_{t+1}~=~(\textbf{I}-\textbf{K}_{t+1}\textbf{J}_{t+1})\hat{\textbf{E}}_{t+1}$.
Fig.~\ref{fig: CDF} depicts the Cumulative Distribution Function (CDF) of 2D positioning error for different positioning techniques. Looking at the CDF curves, one can note the benefits of fusing ISAC and IMU measurements. Our proposed DNN-based positioning technique and baseline~$2$, utilizing the additional source of information from the IMU sensor, outperform the baseline~$1$, which uses only ISAC measurements. Looking at the $90$th percentile, the achievable positioning errors with baseline~$1$, baseline~$2$, and with our proposed approach are $10.5$~cm, $7$~cm, and $5.5$~cm, respectively. Fusing ISAC and IMU measurements with EKF achieves roughly $40\%$ lower positioning errors compared to baseline~$1$. In addition, our proposed two-stage cascaded DNN achieves roughly $62\%$ lower positioning errors compared with the baseline~$1$. On the other hand, comparing the fusion performance between EKF and DNN, our approach achieves roughly $24\%$ lower positioning errors compared with baseline~$2$. Table~\ref{tab:tab1} presents the summary of positioning errors w.r.t. average error and $90$th percentile.

\begin{figure}[t!]
	\centering
	\includegraphics[clip, trim=3.3cm 1cm 4.8cm 5.5cm, width=0.38\textwidth]{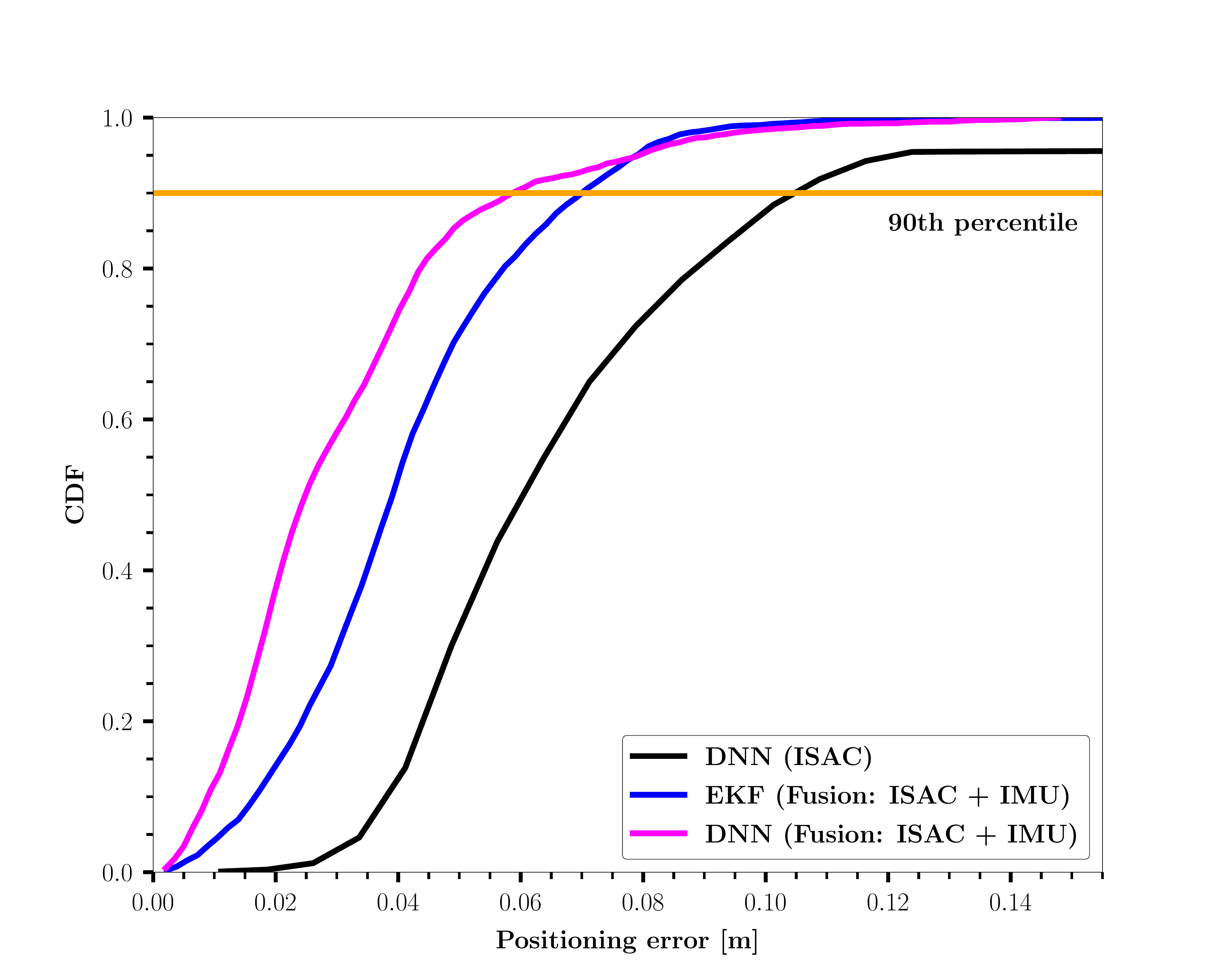}%
	\caption{CDF of 2D positioning error for three positioning techniques.}
	\label{fig: CDF}

\end{figure}


\begin{table}[!t]
    \renewcommand*{\arraystretch}{1.6}
    \centering
    \caption{Summary of 2D positioning error.}
    \begin{tabular}{|p{23mm}|p{15.8mm}|p{15.8mm}|p{16.3mm}|} \hline
         &  {DNN (ISAC)}  &  {EKF (Fusion)}  &  {DNN (Fusion)} \\ \cline{2-4} 
         Average error [cm] & $8.2$ & $5.6$ & $3$  \\ \hline
         $90$th percentile [cm] & $10.5$ & $7$ & $5.5$ \\ \hline
    \end{tabular}
    \label{tab:tab1}
    \squeezeup
    \squeezeup
    
\end{table}
\squeezeup

\section{Conclusion}
This paper presents the positioning results of a target AGV in an indoor scenario using the experimental ISAC PoC built on the existing 5G communications hardware with integrated sensing functionality. The measurements from the non-3GPP sensor IMU available on the target AGV were fused with ISAC sensing information to enhance the ISAC-assisted positioning accuracy. A two-stage cascaded DNN was proposed as the data fusion and positioning solution to achieve this. Our proposed positioning solution avoided more significant errors for the indoor scenario considered in this work. We achieved an average positioning error of $3$~cm, corresponding to around $1\%$ relative positioning accuracy. For $90\%$ of the cases, the achieved positioning error was less than or equal to $5.5$~cm. 

\section*{Acknowledgment}
This work has received funding from the European Union's Horizon 2020 research and innovation programme under the Marie Sklodowska-Curie grant agreement ID~956670. Also, part of this work has been supported by the German Federal Ministry of Education and Research (Foerderkennzeichen 16KISK116, KOMSENS-6G). In addition, this work is part of the project ROUTE56 with grant PID2019-104945GB-I00 funded by MCIN/AEI/ 10.13039/501100011033 and the project 6-SENSES with grant PID2022-138648OB-I00 funded by MCIN/AEI/ 10.13039/501100011033 and by ERDF A way of making Europe. 

\bibliographystyle{IEEEtran}
\bibliography{IEEEabrv,references}
 
\end{document}